\documentclass{article}
\usepackage{amsmath}

\input{tcilatex}

\begin{document}

\title{D-brane worlds and the cosmological constant }
\author{P. Gusin, J. Warczewski \\
University of Silesia, Institute of Physics, ul. Uniwersytecka 4, \\
PL-40007 Katowice, Poland\\
e-mail: pgusin@us.edu.pl }
\date{}
\maketitle

\begin{abstract}
The cosmological constant on a D-brane is analyzed. This D-brane is in the
background produced by the p-brane solutions. The energy-momentum tensor in
this model has been found and the form of the cosmological constant has been
derived. This energy-momentum tensor is interpreted as an energy-momentum
tensor for a perfect fluid on the D-brane. The energy density and the
pressure for this fluid have been derived. As it turned out the pressure is
negative but the speed of sound is real.

\begin{description}
\item[PACS ] 11.25.-w; 11.25.Uv; 98.80.Cq

\item[Keywords] :D-branes, p-branes, cosmological constant
\end{description}
\end{abstract}

\date{}

\section{Introduction}

The problem of the cosmological constant, interpreted as a vacuum energy,
consists in that the vacuum energy obtained from the general relativity (GR)
equations is much smaller, then the vacuum energy obtained from the particle
theory (standard model). This discrepancy can be overcome if one chooses the
initial conditions with the highest accuracy. This leads to the so called
fine-tuning problem. In order to solve these problems several models have
been proposed. One of these consists in the modification of the GR on the
distances bigger then size of the present universe [1]. The extra
dimensions, in this model, have to remain infinite (non-compact) in order to
get consistent theory. From the other side the fine-tuning problem is solved
by the statistical approach to the different vacua of the superstring theory
with the compactified extra dimensions [2]. Each vacuum realizes a
4-dimensional particle theory with a hidden sector. Parameters of this
sector determinate, among others, the vacuum energy of the 4-dimensional
universe. Thus in the huge number of the superstring vacua some part of them
can realize the observed small value of the cosmological constant. The
p-brane solutions of the low-energetic supergravity in the type IIA/IIB
string theory and the discovery of the D-branes in the open string theory
give new view on the cosmological models ([3], [4]). These branes are
extended and interact with each other by gravity. Each brane interacts also
with itself.

We consider an energy-momentum tensor induced on the D-brane by the
non-trivial background given by p-brane solutions. This tensor projected on
the D-brane world-volume has an interpretation of the cosmological constant.
We present an explicit form of the cosmological constant as a function of
the transverse directions to the D-brane.

\section{Gravity generated by $p$-branes}

The form of the gravity when the fundamental constitutes of matter are $p$%
-branes is considered e.g. in [3], [4], [5] and [10].

Let us recall the form of an action and the solutions for the system
consisting of a dilaton $\phi $, a graviton $g_{MN}$\ and an antisymmetric
tensor $A_{M_{1}...M_{d}}$ of arbitrary rank $d$ in a $D$ dimensional
space-time $R^{D}$ coupled to an extended object.

The action $I_{D}\left( d\right) $ for $\phi ,g_{MN},A_{M_{1}...M_{d}}$ has
the form [10]:%
\begin{equation}
I_{D}\left( d\right) =1/2\kappa ^{2}\int_{R^{D}}d^{D}x\sqrt{-g}(R\left(
g\right) -1/2\left\vert d\phi \right\vert ^{2}-\frac{1}{2\left( d+1\right) !}%
e^{-\phi a\left( d\right) }F^{2}),  \tag{2.1}
\end{equation}

where: 
\begin{gather*}
F^{2}=F_{M_{1}...M_{d+1}}F^{M_{1}...M_{d+1}}, \\
F_{M_{1}...M_{d+1}}=\left( dA\right) _{M_{1}...M_{d+1}}.
\end{gather*}%
The above fields are coupled to an elementary $d$-dimensional extended
object ($\left( d-1\right) $-brane) $M$ with a world-volume metric $\gamma
_{\mu \nu }$. This brane is embedded into $R^{D}$: 
\begin{equation*}
X:M\rightarrow R^{D}.
\end{equation*}%
An action $S_{d}$ for this brane is given by: 
\begin{gather}
S_{d}=T_{d}\int_{M\times \mathbf{R}}d^{d}\xi \lbrack -\frac{1}{2}\sqrt{%
-\gamma }\gamma ^{\mu \nu }\partial _{\mu }X^{M}\partial _{\nu
}X^{N}g_{MN}e^{a\phi /d}+  \notag \\
\frac{d-2}{2}\sqrt{-\gamma }  \notag \\
-\frac{1}{d!}\varepsilon ^{\mu _{1}...\mu _{d}}\partial _{\mu
_{1}}X^{M_{1}}...\partial _{\mu _{d}}X^{M_{d}}A_{M_{1}...M_{d}}],  \tag{2.2}
\end{gather}%
where $\mu ,\nu =0,1,...,d-1$. Thus the action $I\left( D,d\right) $ for the
system consists of \ the sum of the actions (2.1) and (2.2): 
\begin{equation}
I\left( D,d\right) =I_{D}\left( d\right) +S_{d}.  \tag{2.3}
\end{equation}%
In the action (2.3) there are five independent fields:

\begin{enumerate}
\item an antisymmetric field $A_{M_{1}...M_{d}}$,

\item a metric $g_{MN}$ on $R^{D}$,

\item a dilaton field $\phi $,

\item a vector field $X$ which makes an embedding of the brane $M$ into $%
R^{D}$,

\item a metric $\gamma _{\mu \nu }$ on $M$.
\end{enumerate}

The equations of motion with the respect the above fields are:

\begin{itemize}
\item The condition $\frac{\delta I\left( D,d\right) }{\delta A}=0$ gives
(the Maxwell equations with the sources): 
\begin{equation}
d\ast \left( e^{-a\phi }F\right) =2\kappa ^{2}\ast J,  \tag{2.4}
\end{equation}%
where the current $J$ is given by: 
\begin{gather}
J^{M_{1}...M_{d}}\left( x\right) =  \notag \\
T_{d}\int_{M\times \mathbf{R}}d^{d}\xi \varepsilon ^{\mu _{1}...\mu
_{d}}\partial _{\mu _{1}}X^{M_{1}}...\partial _{\mu _{d}}X^{M_{d}}\chi , 
\tag{2.5}
\end{gather}%
and $\chi =\delta ^{D}\left( x-X\left( \xi \right) \right) /\sqrt{-g}$

\item The Einstein equations $\frac{\delta I\left( D,d\right) }{\delta g}=0$%
: 
\begin{gather}
R_{MN}-\frac{1}{2}g_{MN}R=\frac{1}{2}\left( \partial _{M}\phi \partial
_{N}\phi -\frac{1}{2}g_{MN}\left\vert d\phi \right\vert ^{2}\right)  \notag
\\
+\kappa ^{2}T_{MN}  \notag \\
+\frac{e^{-a\phi }}{2d!}\left( F_{M}^{M_{1}...M_{d}}F_{NM_{1}...M_{d}}-\frac{%
1}{2\left( d+1\right) }g_{MN}F^{2}\right) ,  \tag{2.6}
\end{gather}%
where $T_{MN}=g_{MA}g_{NB}T^{AB}$ is the energy-momentum tensor of the brane 
$M$ : 
\begin{equation}
T^{AB}\left( x\right) =T_{d}\int_{M\times \mathbf{R}}d^{d}\xi \sqrt{-\gamma }%
\gamma ^{\mu \nu }\partial _{\mu }X^{A}\partial _{\nu }X^{B}e^{a\phi /d}\chi
,  \tag{2.7}
\end{equation}%
and $\chi =\delta ^{D}\left( x-X\left( \xi \right) \right) /\sqrt{-g}$

\item The dilaton equation $\frac{\delta I\left( D,d\right) }{\delta \phi }%
=0 $: 
\begin{gather}
\partial _{M}\left( \sqrt{-g}g^{MN}\partial _{N}\phi \right) +\frac{a\sqrt{-g%
}}{2\left( d+1\right) }e^{-a\phi }F^{2}=  \notag \\
=-\kappa ^{2}\frac{a}{d}T_{d}\int_{M}d^{d}\xi \sqrt{-\gamma }\gamma
^{ij}\partial _{i}X^{M}\partial _{j}X^{N}g_{MN}e^{a\phi /d}\chi \sqrt[-]{-g}.
\tag{2.8}
\end{gather}

\item The brane equations $\frac{\delta I\left( D,d\right) }{\delta X}=0$: 
\begin{gather}
\partial _{\mu }\left( \sqrt{-\gamma }\gamma ^{\mu \nu }\partial _{\nu
}X^{N}g_{MN}e^{a\phi /d}\right) +  \notag \\
-\frac{1}{2}\sqrt{-\gamma }\gamma ^{\mu \nu }\partial _{\mu }X^{N}\partial
_{\nu }X^{P}\partial _{M}\left( g_{NP}e^{a\phi /d}\right) =  \notag \\
\frac{1}{d!}\varepsilon ^{\mu _{1}...\mu _{d}}\partial _{\mu
_{1}}X^{M_{1}}...\partial _{\mu _{d}}X^{M_{d}}F_{MM_{1}...M_{d}}.  \tag{2.9}
\end{gather}

\item The equations of motion $\frac{\delta I\left( D,d\right) }{\delta
\gamma }=0$ for the world metric $\gamma $: 
\begin{equation}
\gamma _{\mu \nu }=\partial _{\mu }X^{M}\partial _{\nu }X^{N}g_{MN}e^{a\phi
/d}.  \tag{2.10}
\end{equation}
\end{itemize}

In order to solve the above coupled system of equations (2.4-2.9), it is
assumed that $R^{D}$ has the topology of the Cartesian product [10]: 
\begin{equation}
R^{D}=M\times N,  \tag{2.11}
\end{equation}%
where $M$ is a $d$-dimensional manifold (($d-1$)-brane) with the Poincare
symmetry group $P\left( d\right) $ and $N$ is an isotropic manifold with $%
SO\left( D-d\right) $ symmetry group.\ The coordinates on $R^{D}$ are split: 
\begin{equation*}
X^{M}=(x^{\mu },y^{m}),
\end{equation*}%
where $x^{\mu }$, $y^{m}$ concerns $M$ and $N,$\ respectively. The indices $%
\mu $ and $m$ have the range:\ $\mu =0,1,...,d-1$ ; $m=1,...,D-d$. In this
topology the ansatz for the metric $g_{MN}$ has the form: 
\begin{equation}
ds^{2}=g_{MN}dx^{M}dx^{N}=e^{2A\left( y\right) }\eta _{\mu \nu }dx^{\mu
}dx^{\nu }+e^{2B\left( y\right) }\delta _{mn}dy^{m}dy^{n},  \tag{2.12}
\end{equation}%
the metric $\eta $ is diagonal: $\left( \eta _{\mu \nu }\right)
=diag(+1,-1,...,-1)$. The functions $A$ and $B$ depend only on $y=\left( 
\mathbf{y\cdot y}\right) ^{1/2}$. The form of the antisymmetric field $A$ is
assumed below: 
\begin{equation}
A_{\mu _{1}...\mu _{d}}=-\frac{1}{\det \left( g_{\mu \nu }\right) }%
\varepsilon _{\mu _{1}...\mu _{d}}e^{C\left( y\right) },  \tag{2.13}
\end{equation}%
where: 
\begin{equation*}
\varepsilon _{\mu _{1}...\mu _{d}}=g_{\mu _{1}\nu _{1}}...g_{\mu _{d}\nu
_{d}}\varepsilon ^{\nu _{1}...\nu _{d}},
\end{equation*}%
($\varepsilon ^{01...d-1}=+1$) and the other components of $A$ are set to
zero, $\det \left( g_{\mu \nu }\right) =\left( -1\right) ^{d-1}e^{2Ad}$.
Thus the field $F$ has the form: 
\begin{equation}
F_{m\mu _{1}...\mu _{d}}=-\frac{1}{\det \left( g_{\mu \nu }\right) }%
\varepsilon _{\mu _{1}...\mu _{d}}\partial _{m}\left( e^{C\left( y\right)
}\right) .  \tag{2.14}
\end{equation}%
The dilaton field $\phi $ depends on $y$ since $N$ is isotropic: 
\begin{equation*}
\phi =\phi \left( y\right) .
\end{equation*}
A static gauge choice for the vector field $X$ is also assumed: 
\begin{equation*}
X^{\mu }=\xi ^{\mu },
\end{equation*}%
where $\xi ^{\mu }$ are coordinates on the brane $M$. In this static gauge
the field $X$ is equal to: 
\begin{equation*}
X^{M}=\left( \xi ^{\mu },Y^{m}\right) .
\end{equation*}%
The directions $Y$ transverse to the brane $M$ are constant: $Y^{m}=const.$
It means that the brane is not moving in this special coordinates system.
Under the above conditions the metric $\gamma $ (Eq. (2.10)) takes the form: 
\begin{equation*}
\gamma _{\mu \nu }=\eta _{\mu \nu }e^{2A+a\phi /d}.
\end{equation*}%
One of the solutions for the above system with the flat asymptotic condition
( $g_{MN}\rightarrow \eta _{MN}$ ) is given by [10]: 
\begin{eqnarray}
A\left( y\right) &=&\frac{\widetilde{d}}{2\left( d+\widetilde{d}\right) }%
\left( C\left( y\right) -C_{0}\right) ,  \TCItag{2.15} \\
B\left( y\right) &=&-\frac{d}{2\left( d+\widetilde{d}\right) }\left( C\left(
y\right) -C_{0}\right) ,  \TCItag{2.16} \\
e^{-C\left( y\right) } &=&\left\{ 
\begin{array}{cc}
e^{-C_{0}}+\frac{k_{d}}{y^{\widetilde{d}}} & \text{for }\widetilde{d}>0 \\ 
e^{-C_{0}}+\frac{\kappa ^{2}T_{d}}{\pi }\ln y & \text{for }\widetilde{d}=0%
\end{array}%
\right. ,  \TCItag{2.17} \\
\frac{a}{d}\phi \left( y\right) &=&\frac{a^{2}}{4}\left( C\left( y\right)
-C_{0}\right) +C_{0},  \TCItag{2.18} \\
a^{2}\left( d\right) &=&4-\frac{2\widetilde{d}d}{d+\widetilde{d}}, 
\TCItag{2.19}
\end{eqnarray}%
where: 
\begin{equation}
k_{d}=\frac{2\kappa ^{2}T_{d}}{\Omega _{\widetilde{d}+1}\widetilde{d}} 
\tag{2.20}
\end{equation}%
and $\widetilde{d}=D-d-2$, $\Omega _{\widetilde{d}+1}$ is the volume of a ($%
\widetilde{d}+1$)-dimensional sphere $S^{\widetilde{d}+1}$. Thus the metric $%
g_{MN}$ is given by: 
\begin{gather}
g_{MN}dX^{M}dX^{N}=\left( 1+\frac{k_{d}}{y^{\widetilde{d}}}e^{-C_{0}}\right)
^{-\frac{\widetilde{d}}{d+\widetilde{d}}}\eta _{\mu \nu }dx^{\mu }dx^{\nu }+
\notag \\
\left( 1+\frac{k_{d}}{y^{\widetilde{d}}}e^{-C_{0}}\right) ^{\frac{d}{d+%
\widetilde{d}}}\delta _{mn}dy^{m}dy^{n}.  \tag{2.21}
\end{gather}

The other solution for this system is given by a ($d+2$)-dimensional
black-brane with the symmetry group: 
\begin{equation*}
\mathbf{R\times }SO\left( d+1\right) \times SO\left( \widetilde{d}-1\right) .
\end{equation*}%
The metric for this system has the form: 
\begin{gather}
ds^{2}=-\Delta _{+}\Delta _{-}^{-\frac{\widetilde{d}}{d+\widetilde{d}}%
}dt^{2}+\Delta _{+}^{-1}\Delta _{-}^{\frac{a^{2}}{2d}-1}dr^{2}+r^{2}\Delta
_{-}^{\frac{a^{2}}{2d}}d\Omega _{d+1}^{2}  \notag \\
+\Delta _{-}^{\frac{d}{d+\widetilde{d}}}dX_{i}dX^{i},  \tag{2.22}
\end{gather}%
where $i=1,...,\widetilde{d}-1$ and: 
\begin{eqnarray}
e^{-2\phi } &=&\Delta _{-}^{a},  \TCItag{2.23} \\
\Delta _{\pm } &=&1-\left( \frac{r_{\pm }}{r}\right) ^{d},  \TCItag{2.24} \\
F_{d+1} &=&\left( r_{+}r_{-}\right) ^{d/2}\varepsilon _{d+1}d,  \TCItag{2.25}
\end{eqnarray}%
and $\varepsilon _{d+1}$ is the volume form of the ($d+1$)-dimensional
sphere $S^{d+1}$with the metric $d\Omega _{d+1}^{2}=h_{ab}d\varphi
^{a}d\varphi ^{b}$. The radii $r_{+}$ and $r_{-}$ are related to the mass $%
\emph{M}_{\widetilde{d}}$ per unit ($\widetilde{d}-1$)-volume and to the
magnetic charge $g_{\widetilde{d}}$: 
\begin{eqnarray}
\emph{M}_{\widetilde{d}} &=&\int d^{D-d}\Theta _{00}=\frac{\Omega _{d+1}}{%
2\kappa ^{2}}[\left( d+1\right) r_{+}^{d}-r_{-}^{d}],  \TCItag{2.26} \\
g_{\widetilde{d}} &=&\frac{1}{\sqrt{2}\kappa }\int_{S^{d+1}}e^{-a\phi }\ast
F=\frac{\Omega _{d+1}}{\sqrt{2}\kappa }d\left( r_{+}r_{-}\right) ^{d/2}, 
\TCItag{2.27}
\end{eqnarray}%
where $\Theta _{MN}$ is the total energy-momentum tensor for the system and $%
\ast $ is the Hodge duality operator with respect to the metric (2.22). In
the case when $r_{+}=r_{-}=r_{0}$ the mass and charge are given by: 
\begin{equation}
\emph{M}_{\widetilde{d}}=\sqrt{2}\kappa g_{\widetilde{d}}.  \tag{2.28}
\end{equation}%
It means that this brane becomes the extremal p-brane (BPS state). In this
case the metric (2.22) takes the form: 
\begin{equation}
ds^{2}=\Delta ^{\frac{d}{d+\widetilde{d}}}\left( -dt^{2}+dX_{i}dX^{i}\right)
+\Delta ^{-\frac{\widetilde{d}}{d+\widetilde{d}}}\left( d\rho ^{2}+\rho
^{2}d\Omega _{d+1}^{2}\right) ,  \tag{2.29}
\end{equation}%
where $\rho ^{d}=r^{d}-r_{0}^{d}$ and $\Delta =1+\left( r_{0}/\rho \right)
^{d}$.

\section{A D-brane motion in the field of the black-brane}

We consider a $D_{d^{\prime }-1}$-brane $M$ embedded in the background of
the ($d+2$)-blackbrane $N$ in the $D$-dimensional space-time $R^{D}$. This ($%
d+2$)-blackbrane wraps a $\left( d+1\right) $-dimensional sphere. The metric
of $R^{D}$ in the presence of blackbrane is given by Eq. (2.22). Thus the
metric $\gamma _{\alpha \beta }$ induced on $M$ by $g_{MN}$ has the form: 
\begin{equation}
\gamma _{\alpha \beta }=g_{MN}\frac{\partial X^{M}}{\partial \xi ^{\alpha }}%
\frac{\partial X^{N}}{\partial \xi ^{\beta }},  \tag{3.1}
\end{equation}%
where $X$ is an embedding of $M$ in $R^{D}$: 
\begin{eqnarray*}
X &:&N\times \mathbf{R}^{1}\rightarrow R^{D}, \\
X^{M} &=&X^{M}\left( \xi ^{0},\xi ^{a}\right)
\end{eqnarray*}%
and $\alpha ,\beta =0,1,...,d^{\prime }-1$ , $n=1,...,d^{\prime }-1$. We
assume that the time in $R^{D}$ and in the worldvolume $M$ is the same and $%
d^{\prime }-1$ directions of $N$ are parallel to $M.$ Thus the embedding of $%
X^{M}$ has the form: 
\begin{equation}
X^{M}\left( \xi ^{0},\xi ^{a}\right) =\left( \xi ^{0},\xi ^{a},X^{m}\left(
\xi ^{0}\right) \right) ,  \tag{3.2}
\end{equation}%
where $a=1,...,d^{\prime }-1$ and $m=1,...,D-d^{\prime }$. The coordinates
on $M$ and $R^{D}$ selected in this way form the static gauge. For the
metric $g_{MN}$ (which is produced by (d+2)-brane wrapped on $S^{d+1}$)
equal to%
\begin{equation}
ds^{2}=\lambda _{0}dt^{2}+\lambda _{1}\sum_{i=1}^{\widetilde{d}%
-1}dX_{i}^{2}+\lambda _{2}dr^{2}+r^{2}\lambda _{3}d\Omega _{d+1}  \tag{3.3}
\end{equation}%
the metric $\gamma _{\alpha \beta }$ induced by the embedding (3.2) takes
the form:%
\begin{eqnarray}
\gamma _{00} &=&\lambda _{0}+\lambda _{1}\sum_{i=d^{\prime }}^{\widetilde{d}%
-1}\overset{\cdot }{X}_{i}^{2}+\lambda _{2}\overset{\cdot }{r}%
^{2}+r^{2}\lambda _{3}\overset{\cdot }{\mathbf{\varphi }}^{2},  \TCItag{3.4}
\\
\gamma _{ab} &=&\lambda _{1}\delta _{ab},\text{for }d^{\prime }-1\leq 
\widetilde{d}-1  \TCItag{3.5}
\end{eqnarray}%
and $\gamma _{a0}=0$, where:%
\begin{equation*}
\overset{\cdot }{\mathbf{\varphi }}^{2}=h_{rs}\overset{\cdot }{\varphi }^{r}%
\overset{\cdot }{\varphi }^{s},
\end{equation*}%
and $h_{rs}=h_{rs}\left( \varphi \right) $ ($r,s=1,...,d+1$)\ is the metric
on $S^{d+1}$. The coordinates $X^{m}$ in the metric (3.3) are as follows:%
\begin{equation*}
X^{m}=\left( X^{i},r,\varphi ^{s}\right) ,
\end{equation*}%
where $i=d^{\prime },...,\widetilde{d}-1$. In the case when the metric of
the background has the form:%
\begin{equation}
ds^{2}=\lambda _{0}dt^{2}+\lambda _{1}\sum_{i=1}^{\widetilde{d}%
-1}dX_{i}^{2}+\lambda _{2}\sum_{m=1}^{d+2}dX_{m}^{2},  \tag{3.6}
\end{equation}%
the induced metric takes the form:%
\begin{eqnarray}
\gamma _{00} &=&\lambda _{0}+\lambda _{1}\sum_{i=d^{\prime }}^{\widetilde{d}%
-1}\overset{\cdot }{X}_{i}^{2}+\lambda _{2}\sum_{m=1}^{d+2}\overset{\cdot }{X%
}_{m}^{2},  \TCItag{3.7} \\
\gamma _{0a} &=&0,  \TCItag{3.8} \\
\gamma _{ab} &=&\lambda _{1}\delta _{ab}\text{, for }d^{\prime }-1\leq 
\widetilde{d}-1.  \TCItag{3.9}
\end{eqnarray}%
If $d^{\prime }-1\geq \widetilde{d}-1$ the metric $\gamma $ is given by:%
\begin{eqnarray}
\gamma _{00} &=&\lambda _{0}+\lambda _{2}\sum_{m=1}^{d+2}\overset{\cdot }{X}%
_{m}^{2},  \TCItag{3.10} \\
\gamma _{a_{1}b_{1}} &=&\lambda _{1}\delta _{a_{1}b_{1}}\text{ for }%
a_{1},b_{1}=1,...,\widetilde{d}-1,  \TCItag{3.11a} \\
\gamma _{a_{2}b_{2}} &=&\lambda _{2}\delta _{a_{2}b_{2}}\text{ for }%
a_{2},b_{2}=\widetilde{d},...,d^{\prime }-1.  \TCItag{3.11b}
\end{eqnarray}

Thus in the gauge (3.2)\ the metric $\gamma $ induced on $M$ by the
blackbrane $N$ (the latter producing the background metric (2.27)) has the
form: 
\begin{eqnarray}
\gamma _{00} &=&-\Delta _{+}\Delta _{-}^{-\frac{\widetilde{d}}{d+\widetilde{d%
}}}+\Delta _{+}^{-1}\Delta _{-}^{\frac{a^{2}}{2d}-1}\overset{\cdot }{r}%
^{2}+r^{2}\Delta _{-}^{\frac{a^{2}}{2d}}\overset{\cdot }{\mathbf{\varphi }}%
^{2}+\Delta _{-}^{\frac{d}{d+\widetilde{d}}}\overset{\cdot }{X}_{i}\overset{%
\cdot }{X}^{i},  \TCItag{3.12} \\
\gamma _{0a} &=&0,  \TCItag{3.13} \\
\gamma _{ab} &=&\Delta _{-}^{\frac{d}{d+\widetilde{d}}}\delta _{ab}. 
\TCItag{3.14}
\end{eqnarray}%
For the static case $\overset{\cdot }{X}_{i}=0$ ($i=1,...,\widetilde{d}-1$) $%
\gamma _{00}$ takes the form: 
\begin{equation}
\gamma _{00}=-\Delta _{+}\Delta _{-}^{-\frac{\widetilde{d}}{d+\widetilde{d}}%
}+\Delta _{+}^{-1}\Delta _{-}^{\frac{a^{2}}{2d}-1}\overset{\cdot }{r}%
^{2}+r^{2}\Delta _{-}^{\frac{a^{2}}{2d}}\overset{\cdot }{\mathbf{\varphi }}%
^{2}.  \tag{3.15}
\end{equation}

\section{The energy-momentum tensor for a D-brane}

The energy-momentum tensor of the ($d^{\prime }-1$)-brane $M$ in the
background $g_{MN}$ is given by Eq. (2.7) and is expressed by the matrix: 
\begin{equation}
\left( T^{MN}\right) =\left( 
\begin{array}{ccc}
T^{00} & T^{0a} & T^{0m} \\ 
T^{a0} & T^{ab} & T^{am} \\ 
T^{m0} & T^{ma} & T^{mn}%
\end{array}%
\right) ,  \tag{4.1}
\end{equation}%
where $a,b=1,...,d^{\prime }-1$ and $m,n=1,...,D-d^{\prime }$ . The generic
form of the background metric $g_{MN}$ produced by a ($d-1$)-brane is given
by (see (3.3) and (3.6)): 
\begin{equation}
\left( g_{MN}\right) =\left( 
\begin{array}{cc}
\left( 
\begin{array}{cc}
\lambda _{0} & \left( 0\right) \\ 
\left( 0\right) & \lambda _{1}I_{\widetilde{d}-1}%
\end{array}%
\right) & 0 \\ 
0 & \left( g_{rs}\right)%
\end{array}%
\right) ,  \tag{4.2}
\end{equation}%
where $I_{\widetilde{d}-1}$ is ($\widetilde{d}-1$)-dimensional unit matrix
and $r,s=1,...,d+2$. Thus the induced metric $\gamma _{\mu \nu }$ on the
brane $M$ for the embedding (3.2) has the form given either by Eqs.(3.4-3.5)
or by Eqs.(3.7-3.11). The components of the energy-momentum tensor $T^{MN}$
for the metric (4.2) in the embedding (3.2) take the form: 
\begin{eqnarray}
T^{\mu \nu } &=&T_{d}\sqrt{\frac{\gamma }{g}}\gamma ^{\mu \nu }e^{a\phi /d}%
\widehat{\delta },  \TCItag{4.3} \\
T^{m0} &=&T^{0m}=T_{d}\sqrt{\frac{\gamma }{g}}\gamma ^{00}\overset{\cdot }{X}%
^{m}e^{a\phi /d}\widehat{\delta },  \TCItag{4.4} \\
T^{mn} &=&T_{d}\sqrt{\frac{\gamma }{g}}\gamma ^{00}\overset{\cdot }{X}^{m}%
\overset{\cdot }{X}^{n}e^{a\phi /d}\widehat{\delta },  \TCItag{4.5}
\end{eqnarray}%
where $\widehat{\delta }=\delta ^{D-d}\left( x^{m}-X^{m}\left( \xi
^{0}\right) \right) $. Other components ($T^{a0}$ , $T^{am}$) are equal to
zero and $\gamma =\det \left( \gamma _{\mu \nu }\right) $, $g=\det \left(
g_{MN}\right) $.

\subsection{Cosmological constant induced by the blackbranes}

The ratio of the determinants $\gamma /g$ for the metrics (3.3), (3.4) and
(3.5) is given by: 
\begin{equation}
\gamma /g=\frac{\lambda _{1}^{d^{\prime }-\widetilde{d}}\Gamma }{r^{2\left(
d+1\right) }\lambda _{2}\lambda _{3}^{d+1}\det h},  \tag{4.6}
\end{equation}%
where: 
\begin{equation}
\Gamma =1+\frac{\lambda _{1}}{\lambda _{0}}\sum_{i=d^{\prime }}^{\widetilde{d%
}-1}\overset{\cdot }{X}_{i}^{2}+\frac{\lambda _{2}}{\lambda _{0}}\overset{%
\cdot }{r}^{2}+r^{2}\frac{\lambda _{3}}{\lambda _{0}}\overset{\cdot }{%
\mathbf{\varphi }}^{2},  \tag{4.7}
\end{equation}%
and $\det h$ is the determinant of the metric $h_{rs}$ on the sphere $%
S^{d+1} $.\ For the embedding (3.2) and in the metric (3.3) the
time-dependent components $\overset{\cdot }{X}^{m}$in Eqs.(4.4-4.5) have the
form:%
\begin{equation*}
\overset{\cdot }{X}^{m}=\left( \overset{\cdot }{X}^{i},\overset{\cdot }{r},%
\overset{\cdot }{\varphi }^{s}\right) ,
\end{equation*}%
where $i=d^{\prime },...,\widetilde{d}-1$ and $s=1,...,d+1$ ($\varphi ^{s}$
are coordinates on $S^{d+1}$). In this way we obtain the explicit form of $%
T^{MN}$: 
\begin{eqnarray}
T^{\mu \nu } &=&\frac{T_{d^{\prime }}}{r^{d+1}}\sqrt{\frac{\lambda
_{1}^{d^{\prime }-\widetilde{d}}\Gamma }{\lambda _{2}\lambda _{3}^{d+1}\det h%
}}\gamma ^{\mu \nu }e^{a\phi /d}\widehat{\delta },  \TCItag{4.8} \\
T^{m0} &=&\frac{T_{d^{\prime }}}{r^{d+1}}\sqrt{\frac{\lambda _{1}^{d^{\prime
}-\widetilde{d}}\Gamma }{\lambda _{2}\lambda _{3}^{d+1}\det h}}\frac{\overset%
{\cdot }{X}^{m}e^{a\phi /d}}{\lambda _{0}\Gamma }\widehat{\delta }, 
\TCItag{4.9} \\
T^{mn} &=&\frac{T_{d^{\prime }}}{r^{d+1}}\sqrt{\frac{\lambda _{1}^{d^{\prime
}-\widetilde{d}}\Gamma }{\lambda _{2}\lambda _{3}^{d+1}\det h}}\frac{\overset%
{\cdot }{X}^{m}\overset{\cdot }{X}^{n}e^{a\phi /d}}{\lambda _{0}\Gamma }%
\widehat{\delta },  \TCItag{4.10}
\end{eqnarray}%
since $\gamma _{00}=\lambda _{0}\Gamma $ and $\gamma ^{00}=\left( \lambda
_{0}\Gamma \right) ^{-1}$. The $D$-brane tension $T_{d^{\prime }}$ is given
by [12, 13]: 
\begin{equation*}
T_{d^{\prime }}^{2}=\frac{\pi }{\kappa _{\left( 10\right) }^{2}}\left( 4\pi
^{2}\alpha ^{\prime }\right) ^{4-d^{\prime }}.
\end{equation*}%
The pull-back of the $T^{MN}$ by the embedding $X$ gives the energy-momentum
tensor $\widetilde{T}_{\mu \nu }$ on the D-brane: 
\begin{equation}
\widetilde{T}_{\mu \nu }=T^{AB}g_{AM}g_{BN}\frac{\partial X^{M}}{\partial
\xi ^{\mu }}\frac{\partial X^{N}}{\partial \xi ^{\nu }}.  \tag{4.11}
\end{equation}%
Thus we obtain from (4.11): 
\begin{gather*}
\widetilde{T}_{00}=T^{00}g_{00}^{2}+T^{m_{1}n_{1}}g_{m_{1}m}g_{n_{1}n}%
\overset{\cdot }{X}^{m}\overset{\cdot }{X}^{n} \\
\widetilde{T}_{0a}=0, \\
\widetilde{T}_{ab}=T^{cd}g_{ca}g_{bd},
\end{gather*}%
where 
\begin{eqnarray*}
g_{00} &=&\lambda _{0,} \\
\left( g_{m_{1}m}\right) &=&\left( 
\begin{array}{ccc}
\lambda _{1}I_{\widetilde{d}-d^{\prime }} & 0 & 0 \\ 
0 & \lambda _{2} & 0 \\ 
0 & 0 & r^{2}\lambda _{3}\left( h_{ps}\right)%
\end{array}%
\right) , \\
\left( g_{ac}\right) &=&\lambda _{1}I_{d^{\prime }-1}.
\end{eqnarray*}%
Because $g_{mn}\overset{\cdot }{X}^{m}\overset{\cdot }{X}^{n}=\lambda
_{0}\left( \Gamma -1\right) $ we get from (4.8-10): 
\begin{gather}
\widetilde{T}_{00}=\frac{T_{d^{\prime }}}{r^{d+1}}\sqrt{\frac{\lambda
_{1}^{d^{\prime }-\widetilde{d}}}{\lambda _{2}\lambda _{3}^{d+1}\Gamma \det h%
}}e^{a\phi /d}\lambda _{0}\left[ \Gamma ^{2}-2\Gamma +2\right] ,  \tag{4.12}
\\
\widetilde{T}_{ab}=\frac{T_{d^{\prime }}}{r^{d+1}}\sqrt{\frac{\lambda
_{1}^{d^{\prime }-\widetilde{d}}\Gamma }{\lambda _{2}\lambda _{3}^{d+1}\det h%
}}e^{a\phi /d}\lambda _{1}\delta _{ab},  \tag{4.13}
\end{gather}%
modulo delta functions. In the static case ($\overset{\cdot }{X}^{m}=0$) $\
\Gamma =1,$ so the Eqs. (4.12-13) take the form: 
\begin{equation}
\widetilde{T}_{\mu \nu }=\frac{T_{d^{\prime }}}{r^{d+1}}\sqrt{\frac{\lambda
_{1}^{d^{\prime }-\widetilde{d}}}{\lambda _{2}\lambda _{3}^{d+1}\det h}}%
e^{a\phi /d}\gamma _{\mu \nu }.  \tag{4.14}
\end{equation}%
This tensor consists of the part 
\begin{equation}
\Lambda _{b}\left( r;d^{\prime },d\right) =\frac{T_{d^{\prime }}}{r^{d+1}}%
\sqrt{\frac{\lambda _{1}^{d^{\prime }-\widetilde{d}}}{\lambda _{2}\lambda
_{3}^{d+1}\det h}}e^{a\phi /d},  \tag{4.15}
\end{equation}%
which depends only on the direction $r$ transverse to the brane $M$. Thus
for a fixed position of $M$ in the ambient space $R$ the quantity $\Lambda
_{b}$ has the constant value. The equations of gravity on $M$ take the form: 
\begin{equation}
R_{\mu \nu }-\frac{1}{2}\gamma _{\mu \nu }R=t_{\mu \nu }+\widetilde{T}_{\mu
\nu },  \tag{4.16}
\end{equation}%
where $R_{\mu \nu }$ is Ricci tensor and $R$ is scalar curvature with
respect to the metric $\gamma _{\mu \nu }$ and $t_{\mu \nu }$ is the
energy-momentum tensor for the matter and fields on the D-brane. Because $%
\widetilde{T}_{\mu \nu }$ is the product of $\Lambda _{b}$ (which is
constant on $M$) and the metric $\gamma _{\mu \nu },$ the equation (4.16)
takes the form:%
\begin{equation*}
R_{\mu \nu }-\frac{1}{2}\gamma _{\mu \nu }R=t_{\mu \nu }+\Lambda _{b}\gamma
_{\mu \nu }.
\end{equation*}%
Thus $\Lambda _{b}$ can be identified as a cosmological constant which is
produced by the other branes. For the metric (2.22) one obtains that: 
\begin{equation}
\Lambda _{b}\left( r;d^{\prime },d\right) =\frac{T_{d^{\prime }}}{r^{2\left(
d+1\right) }}\sqrt{\frac{\Delta _{+}}{\det h}}\Delta _{-}^{\sigma }, 
\tag{4.17}
\end{equation}%
where:%
\begin{equation}
\sigma \left( d^{\prime },d;D\right) =\frac{\widetilde{d}\left( 3-d\right)
+d^{\prime }d}{2\left( D-2\right) }-\frac{3}{d}+\frac{1}{2}.  \tag{4.18}
\end{equation}%
In the static case induced by the ($d-1)$-dimensional blackbrane the term $%
\Lambda _{b}$ on the $D_{d^{\prime }-1}$-brane takes the form:%
\begin{equation}
\Lambda _{b}\left( r;d^{\prime },d\right) =\frac{T_{d^{\prime }}}{r^{2\left(
d+1\right) }}\sqrt{\frac{1}{\det h}}\left( 1-\frac{r_{+}^{d}}{r^{d}}\right)
^{1/2}\left( 1-\frac{r_{-}^{d}}{r^{d}}\right) ^{\sigma },  \tag{4.19}
\end{equation}%
where the radial coordinate $r$ is interpreted as a distance from the center
of the blackbrane wrapped on $S^{d+1}$ to the center of the ($d^{\prime
}-1)- $brane. The dimensions of the blackbranes change from $0$ to $D-1$.
Thus the total term $\Lambda _{b}$ induced by the set of blackbranes of
different dimensions can be expressed by the following sum:%
\begin{equation}
\Lambda _{b}\left( r_{1},...,r_{D-1};d^{\prime }\right)
=\sum_{d=1}^{D-1}\Lambda _{b}\left( r_{d};d^{\prime },d\right) ,  \tag{4.20}
\end{equation}%
where $r_{d}$ is the distance from ($d-1)-$dimensional brane to ($d^{\prime
}-1)-$brane.

In the case, when $D=10$ and $d^{\prime }=4,$%
\begin{equation}
\sigma \left( 4,d;10\right) =2-\frac{3}{d}+\frac{1}{16}\left( d-7\right) d. 
\tag{4.21}
\end{equation}%
Thus%
\begin{equation}
\Lambda _{b}\left( r;4,d\right) =\frac{T_{4}}{r^{2\left( d+1\right) }}\sqrt{%
\frac{1}{\det h}}\left( 1-\frac{r_{+}^{d}}{r^{d}}\right) ^{1/2}\left( 1-%
\frac{r_{-}^{d}}{r^{d}}\right) ^{2-\frac{3}{d}+\frac{1}{16}\left( d-7\right)
d}  \tag{4.22}
\end{equation}%
and the total cosmological constant is given by:%
\begin{equation}
\Lambda _{b}\left( r_{1},...,r_{9};4\right) =\sum_{d=1}^{9}\Lambda
_{b}\left( r_{d};4,d\right) .  \tag{4.23}
\end{equation}%
In this way we showed that the induced cosmological constant is the function
of dimensions of the blackbranes and distances from them to the
4-dimensional brane.

In the non-static case ($\Gamma \neq 1$) we introduce a scalar field $\phi $
which is related to the\ transverse coordinates of the blackbrane:%
\begin{equation}
\phi ^{2}=\frac{\lambda _{1}}{\left\vert \lambda _{0}\right\vert }%
\sum_{i=d^{\prime }}^{\widetilde{d}-1}\overset{\cdot }{X}_{i}^{2}+\frac{%
\lambda _{2}}{\left\vert \lambda _{0}\right\vert }\overset{\cdot }{r}%
^{2}+r^{2}\frac{\lambda _{3}}{\left\vert \lambda _{0}\right\vert }\overset{%
\cdot }{\mathbf{\varphi }}^{2}.  \tag{4.24}
\end{equation}%
Thus:%
\begin{equation}
\Gamma =1-\phi ^{2},  \tag{4.25}
\end{equation}%
since $\lambda _{0}$ is negative. The Eqs. (4.12-13) take the forms:%
\begin{eqnarray}
\widetilde{T}_{00} &=&\Lambda _{b}\frac{1+\phi ^{4}}{\sqrt{1-\phi ^{2}}}%
\gamma _{00},  \TCItag{4.26} \\
\widetilde{T}_{ab} &=&\Lambda _{b}\sqrt{1-\phi ^{2}}\gamma _{ab}, 
\TCItag{4.27}
\end{eqnarray}%
where $\Lambda _{b}$ is given by (4.19) or by (4.22) for $d^{\prime }=4$.
Let us compare in the commoving frame ($u_{a}=0$) this induced
energy-momentum tensor to an energy-momentum tensor for a perfect fluid $%
T_{\mu \nu }$ with an energy density $\varepsilon $ and a pressure $p$:%
\begin{equation}
T_{\mu \nu }=\left( \varepsilon +p\right) u_{\mu }u_{\nu }-p\gamma _{\mu \nu
}.  \tag{4.28}
\end{equation}
As a result of this comparison one obtains:%
\begin{eqnarray}
\varepsilon &=&\Lambda _{b}\frac{1+\phi ^{4}}{\sqrt{1-\phi ^{2}}}, 
\TCItag{4.29} \\
p &=&-\Lambda _{b}\sqrt{1-\phi ^{2}}.  \TCItag{4.30}
\end{eqnarray}%
The corresponding state equation has the form:%
\begin{equation}
w=p/\varepsilon =-\frac{1-\phi ^{2}}{1+\phi ^{4}}.  \tag{4.31}
\end{equation}

For the variety of the blackbranes we get a set of the fields $\phi _{d}.$
Thus the effective energy and the pressure have the form:%
\begin{eqnarray}
\varepsilon &=&\sum_{d=1}^{9}\Lambda _{b}\left( r_{d};4,d\right) \frac{%
1+\phi _{d}^{4}}{\sqrt{1-\phi _{d}^{2}}},  \TCItag{4.32} \\
p &=&-\sum_{d=1}^{9}\Lambda _{b}\left( r_{d};4,d\right) \sqrt{1-\phi _{d}^{2}%
}.  \TCItag{4.33}
\end{eqnarray}%
In this case the state equation is:%
\begin{equation}
w=-\frac{\sum_{d=1}^{9}\Lambda _{b}\left( r_{d};4,d\right) \sqrt{1-\phi
_{d}^{2}}}{\sum_{d=1}^{9}\Lambda _{b}\left( r_{d};4,d\right) \frac{1+\phi
_{d}^{4}}{\sqrt{1-\phi _{d}^{2}}}}.  \tag{4.34}
\end{equation}%
One can see from above that for the certain values of the fields $\phi _{d}$
the state equation assumes the form $w\leq -1/3$ which corresponds to the
exotic matter interpretation on the D$3$-brane.

\subsection{Cosmological constant induced by the branes without horizon}

In this case the background metric is given by (2.21) and the induced metric 
$\gamma $ is given by (3.7-9) for $d^{\prime }-1\leq \widetilde{d}-1$ and by
(3.10-11a,b) for $d^{\prime }-1\geq \widetilde{d}-1$. Thus the ratio of the
corresponding determinants has the form:%
\begin{equation}
\frac{\det \gamma }{\det g}=\left\{ 
\begin{array}{c}
\frac{\lambda _{1}^{d^{\prime }-\widetilde{d}}}{\lambda _{2}^{d+2}}\Gamma 
\text{ for }d^{\prime }-1\leq \widetilde{d}-1, \\ 
\lambda _{2}^{d^{\prime }+2d+4-D}\Omega \text{ for }d^{\prime }-1\geq 
\widetilde{d}-1,%
\end{array}%
\right.  \tag{4.35}
\end{equation}%
where:%
\begin{eqnarray}
\Gamma &=&1+\frac{\lambda _{1}}{\lambda _{0}}\sum_{i=d^{\prime }}^{%
\widetilde{d}-1}\overset{\cdot }{y}_{i}^{2}+\frac{\lambda _{2}}{\lambda _{0}}%
\sum_{m=1}^{d+2}\overset{\cdot }{y}_{m}^{2},  \TCItag{4.36} \\
\Omega &=&1+\frac{\lambda _{2}}{\lambda _{0}}\sum_{m=1}^{d+2}\overset{\cdot }%
{y}_{m}^{2}.  \TCItag{4.37}
\end{eqnarray}%
Let us consider first the case when $d^{\prime }-1\leq \widetilde{d}-1$.
Thus:%
\begin{eqnarray}
T^{\mu \nu } &=&T_{d^{\prime }}\sqrt{\frac{\lambda _{1}^{d^{\prime }-%
\widetilde{d}}\Gamma }{\lambda _{2}^{d+1}}}\gamma ^{\mu \nu }e^{a\phi /d}%
\widehat{\delta },  \TCItag{4.38} \\
T^{m0} &=&T_{d^{\prime }}\sqrt{\frac{\lambda _{1}^{d^{\prime }-\widetilde{d}%
}\Gamma }{\lambda _{2}^{d+1}}}\gamma ^{00}\overset{\cdot }{y}^{m}e^{a\phi /d}%
\widehat{\delta },  \TCItag{4.39} \\
T^{mn} &=&T_{d^{\prime }}\sqrt{\frac{\lambda _{1}^{d^{\prime }-\widetilde{d}%
}\Gamma }{\lambda _{2}^{d+1}}}\gamma ^{00}\overset{\cdot }{y}^{m}\overset{%
\cdot }{y}^{n}e^{a\phi /d}\widehat{\delta }.  \TCItag{4.40}
\end{eqnarray}%
Proceeding as before one obtains the following pull-back of this tensor on
the $D_{d^{\prime }-1}$-brane:%
\begin{gather}
\widetilde{T}_{00}=T^{00}g_{00}^{2}+T^{m_{1}n_{1}}g_{m_{1}m}g_{n_{1}n}%
\overset{\cdot }{y}^{m}\overset{\cdot }{y}^{n}  \tag{4.41} \\
\widetilde{T}_{0a}=0,  \tag{4.42} \\
\widetilde{T}_{ab}=T^{cd}g_{ca}g_{bd},  \tag{4.43}
\end{gather}%
where:\ \ 
\begin{eqnarray}
g_{00} &=&\lambda _{0,}  \TCItag{4.44} \\
\left( g_{m_{1}m}\right) &=&\left( 
\begin{array}{cc}
\lambda _{1}I_{\widetilde{d}-d^{\prime }} &  \\ 
& \lambda _{2}I_{d+2}%
\end{array}%
\right) ,  \TCItag{4.45} \\
\left( g_{ac}\right) &=&\lambda _{1}I_{d^{\prime }-1}.  \TCItag{4.46}
\end{eqnarray}%
Thus:%
\begin{eqnarray}
\widetilde{T}_{00} &=&T_{d^{\prime }}\sqrt{\frac{\lambda _{1}^{d^{\prime }-%
\widetilde{d}}}{\lambda _{2}^{d+1}\Gamma }}e^{a\phi /d}\lambda _{0}\left[
\Gamma ^{2}-2\Gamma +2\right] ,  \TCItag{4.47} \\
\widetilde{T}_{ab} &=&T_{d^{\prime }}\sqrt{\frac{\lambda _{1}^{d^{\prime }-%
\widetilde{d}}\Gamma }{\lambda _{2}^{d+1}}}e^{a\phi /d}\lambda _{1}\delta
_{ab}.  \TCItag{4.48}
\end{eqnarray}%
In the static case ($\Gamma =1$) this energy-momentum tensor has the form:%
\begin{equation}
\widetilde{T}_{\mu \nu }=\Lambda _{o}^{\prime }\gamma _{\mu \nu }, 
\tag{4.49}
\end{equation}%
where:%
\begin{equation}
\Lambda _{o}^{\prime }\left( d^{\prime },d,D;y\right) =T_{d^{\prime }}\left(
1+\frac{k_{d}}{y^{\widetilde{d}}}e^{-C_{0}}\right) ^{\sigma }  \tag{4.50}
\end{equation}%
and 
\begin{equation}
\sigma =\frac{d}{2\left( d+\widetilde{d}\right) }\left( d^{\prime }+2%
\widetilde{d}-2\right) .  \tag{4.51}
\end{equation}%
In the superstring regime $d+\widetilde{d}=8$ and for $d^{\prime }=4$ the
exponent $\sigma $ is equal to:%
\begin{equation}
\sigma =\frac{d\left( 18-d\right) }{16}  \tag{4.52}
\end{equation}%
and $d\leq 4$. Thus the term $\Lambda _{o}^{\prime }$ takes the form:%
\begin{equation}
\Lambda _{o}^{\prime }\left( 4,d,10;y\right) =T_{4}\left( 1+\frac{k_{d}}{y^{%
\widetilde{d}}}e^{-C_{0}}\right) ^{\frac{d\left( 18-d\right) }{16}} 
\tag{4.53}
\end{equation}%
and for $y\rightarrow \infty $ tends to $T_{4}$.

The second case is for $d^{\prime }-1\geq \widetilde{d}-1$. The induced
energy-momentum tensor is given by:%
\begin{eqnarray}
\widetilde{T}_{00} &=&T_{d^{\prime }}\sqrt{\frac{\lambda _{2}^{d^{\prime
}+2d+4-D}}{\Omega }}e^{a\phi /d}\lambda _{0}\left[ \Omega ^{2}-2\Omega +2%
\right] ,  \TCItag{4.54} \\
\widetilde{T}_{ab} &=&T_{d^{\prime }}\sqrt{\lambda _{2}^{d^{\prime
}+2d+4-D}\Omega }e^{a\phi /d}\lambda _{1}\delta _{ab}.  \TCItag{4.55}
\end{eqnarray}%
Thus as before in the static case ($\Omega =1$) we obtain:%
\begin{equation}
\widetilde{T}_{\mu \nu }=\Lambda _{o}^{\prime \prime }\gamma _{\mu \nu }, 
\tag{4.56}
\end{equation}%
where:%
\begin{equation}
\Lambda _{o}^{\prime \prime }\left( d^{\prime },d,D;y\right) =T_{d^{\prime
}}\left( 1+\frac{k_{d}}{y^{\widetilde{d}}}e^{-C_{0}}\right) ^{-\sigma }, 
\tag{4.57}
\end{equation}%
and%
\begin{equation}
\sigma =\frac{\widetilde{d}}{2\left( d+\widetilde{d}\right) }\left(
D+d^{\prime }+d\right) .  \tag{4.58}
\end{equation}%
In the superstring regime $d+\widetilde{d}=8$ and for $d^{\prime }=4$ the
exponent $\sigma $ is equal to:%
\begin{equation}
\sigma =\frac{\left( d+14\right) \left( 10-d\right) }{16}  \tag{4.59}
\end{equation}%
and $d>4$ . Thus the term $\Lambda _{o}^{\prime \prime }$ takes the form:%
\begin{equation}
\Lambda _{o}^{\prime \prime }\left( 4,d,10;y\right) =T_{4}\left( 1+\frac{%
k_{d}}{y^{\widetilde{d}}}e^{-C_{0}}\right) ^{-\frac{\left( d+14\right)
\left( 10-d\right) }{16}}.  \tag{4.60}
\end{equation}

In the presence of the variety of branes with different dimensions the total
induced cosmological constant is given by:%
\begin{equation}
\Lambda \left( y_{1},...,y_{9}\right) =\sum\limits_{d=1}^{4}\Lambda
_{o}^{\prime }\left( 4,d,10;y_{d}\right) +\sum_{d=5}^{9}\Lambda _{o}^{\prime
\prime }\left( 4,d,10;y_{d}\right) .  \tag{4.61}
\end{equation}

In the non-static case we introduce as before a field $\phi $ defined by:%
\begin{equation}
\phi ^{2}=\frac{\lambda _{1}}{\left\vert \lambda _{0}\right\vert }%
\sum_{i=d^{\prime }}^{\widetilde{d}-1}\overset{\cdot }{y}_{i}^{2}+\frac{%
\lambda _{2}}{\left\vert \lambda _{0}\right\vert }\sum_{m=1}^{d+2}\overset{%
\cdot }{y}_{m}^{2}.  \tag{4.62}
\end{equation}%
Thus:%
\begin{eqnarray}
\widetilde{T}_{00} &=&\Lambda \frac{1+\phi ^{4}}{\sqrt{1-\phi ^{2}}}\gamma
_{00},  \TCItag{4.63} \\
\widetilde{T}_{ab} &=&\Lambda \sqrt{1-\phi ^{2}}\gamma _{ab},  \TCItag{4.64}
\end{eqnarray}%
where $\Lambda $ is expressed by (4.53) for $d\leq 4$ and by (4.60) for $d>4$%
. Proceeding as at the end of the previous section one gets energy, pressure
and the state equation for the exotic matter induced by a non-blackbrane on D%
$3$-brane:%
\begin{eqnarray}
\varepsilon &=&\Lambda \frac{1+\phi ^{4}}{\sqrt{1-\phi ^{2}}},  \TCItag{4.65}
\\
p &=&-\Lambda \sqrt{1-\phi ^{2}},  \TCItag{4.66}
\end{eqnarray}%
\begin{equation}
w=p/\varepsilon =-\frac{1-\phi ^{2}}{1+\phi ^{4}},  \tag{4.67}
\end{equation}%
where $\Lambda $ is the same as in (4.63). For the variety of branes one
gets:%
\begin{eqnarray}
\varepsilon &=&\sum_{d=1}^{4}\Lambda _{0}^{\prime }\left( r_{d};4,d\right) 
\frac{1+\phi _{d}^{4}}{\sqrt{1-\phi _{d}^{2}}}+\sum_{d=5}^{9}\Lambda
_{0}^{\prime \prime }\left( r_{d};4,d\right) \frac{1+\phi _{d}^{4}}{\sqrt{%
1-\phi _{d}^{2}}},  \TCItag{4.68} \\
p &=&-\sum_{d=1}^{4}\Lambda _{o}^{\prime }\left( r_{d};4,d\right) \sqrt{%
1-\phi _{d}^{2}}-\sum_{d=1}^{9}\Lambda _{o}^{\prime \prime }\left(
r_{d};4,d\right) \sqrt{1-\phi _{d}^{2}}.  \TCItag{4.69}
\end{eqnarray}%
In this case the state equation is:%
\begin{equation}
w=-\frac{\sum_{d=1}^{4}\Lambda _{o}^{\prime }\left( r_{d};4,d\right) \sqrt{%
1-\phi _{d}^{2}}+\sum_{d=1}^{9}\Lambda _{o}^{\prime \prime }\left(
r_{d};4,d\right) \sqrt{1-\phi _{d}^{2}}}{\sum_{d=1}^{4}\Lambda _{0}^{\prime
}\left( r_{d};4,d\right) \frac{1+\phi _{d}^{4}}{\sqrt{1-\phi _{d}^{2}}}%
+\sum_{d=5}^{9}\Lambda _{0}^{\prime \prime }\left( r_{d};4,d\right) \frac{%
1+\phi _{d}^{4}}{\sqrt{1-\phi _{d}^{2}}}}.  \tag{4.70}
\end{equation}

\subsection{The cosmological constant}

In order to get the total induced cosmological constant on the D3-brane we
should take into account all kinds of p-branes with different dimensions and
with different distances from their centers to the center of D3-brane. Thus
collecting the results (4.23) and (4.61) one obtains:%
\begin{equation}
\Lambda =\sum_{d=1}^{9}\Lambda _{b}\left( 4,d,10;r_{d}\right)
+\sum\limits_{d=1}^{4}\Lambda _{o}^{\prime }\left( 4,d,10;y_{d}\right)
+\sum_{d=5}^{9}\Lambda _{o}^{\prime \prime }\left( 4,d,10;y_{d}\right) . 
\tag{4.71}
\end{equation}%
The dynamic case is described on the D$3$-brane by a set of fields $\{\phi
_{S}\}$ (where $S=0,1,...,9$ is the dimension of the p-brane in the
10-dimensional ambient space).

The global energy-momentum tensor for the perfect fluid on the D$3$-brane is
produced by the configurations of branes of different kinds in
10-dimensional ambient time-space. Thus the cosmological constant induced on
the D3-brane can be fitted to the observed one by the appropriate choice of
the parameters appearing in (4.71). As it is well-known the evolution of the
universe depends on the value of the cosmological constant. In the present
time the universe is accelerated. This phenomenon is being usually explained
by an assumption of the existence of the exotic matter described by the
state equation $w<-1/3.$ Such an exotic matter produces negative pressure,
which acts against gravitation. Our approach presented above allows us to
obtain such a state equation by the appropriate choice of the values of
certain parameters in (4.71). The Eq. (4.71) changes with time which means
that the evolution of D3-brane presented above depends not only on the
D3-brane contents but also on the p-branes configuration in the ambient
space.

\section{Conclusions}

The form of the cosmological constant $\Lambda $ on the D3-brane $M$ has
been derived as a pull-back of the energy-momentum tensor, the latter tensor
being taken for the background produced by the different p-branes. The
contributions coming from the gravity solutions for the p-branes have been
taken into account only and both the gauge fields on $M$ and RR charges of
the p-branes have been ignored. In this way the dependence of the
cosmological constant on both the dimensions of the p-branes and their
distances to $M$ has been obtained.

In the dynamic case when $\phi \neq 0$ one obtains the energy-momentum
tensor on $M$ which can be identified with the energy-momentum tensor for
the perfect fluid on $M$. This perfect fluid representing some kind of the
exotic matter has the state equations given either by Eq.(4.34) in the case
of the blackbranes or by Eq. (4.70) in the case of the p-branes without
horizon. The energy and pressure are given by (4.32) and (4.33) for the
first case as well as by (4.68) and (4.69) for the second case. The pressure
produced by this perfect fluid is negative and acts against gravitation. One
can then say that the perfect fluid can present one of the factors
determining the cosmological evolution of $M$ . Thus the evolution of $M$
depends on its position in the ambient space with respect to the other
branes. As one can see the speed of sound $c_{s}$ is real as a function of
the field $\phi $ for $\phi \in \left( -1,1\right) $ (here only one brane is
considered):%
\begin{equation*}
c_{s}=\sqrt{\frac{dp}{d\varepsilon }}=\sqrt{\frac{1-\phi ^{2}}{1+4\phi
^{2}+\phi ^{3}-3\phi ^{4}}},
\end{equation*}%
where $\varepsilon $ is given by the Eq. (4.29) and $p$ is given by the Eq.
(4.30). The picture below shows the $\phi $-dependence of $c_{s}$: \FRAME{%
dtbpFX}{3in}{2.0003in}{0pt}{}{}{Plot}{\special{language "Scientific
Word";type "MAPLEPLOT";width 3in;height 2.0003in;depth 0pt;display
"USEDEF";plot_snapshots TRUE;mustRecompute FALSE;lastEngine "MuPAD";xmin
"-1";xmax "1";xviewmin "-1.002";xviewmax "1.002";yviewmin "-0.001";yviewmax
"1.001";plottype 4;labeloverrides 3;numpoints 49;plotstyle "patch";axesstyle
"normal";xis \TEXUX{v58144};yis \TEXUX{y};var1name \TEXUX{$\phi $};var2name
\TEXUX{$y$};function \TEXUX{$\sqrt{\frac{1-\phi ^{2}}{1+4\phi ^{2}+\phi
^{3}-3\phi ^{4}}}$};linecolor "black";linestyle 1;pointstyle
"point";linethickness 1;lineAttributes "Solid";var1range
"-1,1";num-x-gridlines 49;curveColor "[flat::RGB:0000000000]";curveStyle
"Line";rangeset"X";valid_file "T";tempfilename
'JZHF6700.wmf';tempfile-properties "XPR";}}The model presented above can be
interpreted as one of the models for the explanation of the origin of the
dark energy. In this model the dark energy is generated by the perfect fluid
of the exotic matter\emph{. }In [14] the dark energy is associated with
p-branes in the light-cone parametrization.

\section{References}

[1] G. R. Dvali, G. Gabadadze, M. Porrati, Phys. Lett. \textbf{B485} 208
(2000) (hep-th/0005016); G. R. Dvali, G. Gabadadze, Phys. Rev. \textbf{D63 }%
065007 (2001) (hep-th/0008054)

[2] M. R. Douglas, \textit{Basic results in Vacuum Statistics,}
hep-th/0409207

[3] A. Kehagis, E. Kiritsis, \textit{Mirage Cosmology},\ hep-th/9910174

[4] E. Kiritsis, \textit{D-branes in Standard Model building, Gravity and
Cosmology}, hep-th/0310001

[5] C. Park, S. J. Sin, \textit{p-brane cosmology and phases of Brans-Dicke
theory with matter}, Phys. Rev. \textbf{D57} (1998) 4620-4628

[6] C. P. Bachas, P. Bain, M. B. Green: JHEP 9905 (1999) 011; \textit{%
Curvature terms in D-brane actions and their M-theory origin,} hep-th/9903210

[7] M. R. Douglas, D. Kabat, P. Pouliot, S. H. Shenker, \textit{D-branes and
Short Distances in String Theory}, hep-th/9608024; Nucl. Phys. \textbf{485}
(1997) 85

[8] S. Weinberg, \textit{Gravitation and Cosmology}, John Wiley and Sons,
Inc., New York 1972

[9] A. Besse, \textit{Einstein Manifolds}, Springer Verlag, Berlin,
Heidelberg, 1987

[10] M. J. Duff, R. R. Khuri, J. X. Lu, \textit{String solitons},
hep-th/9412184 ; M. J. Duff, \textit{Supermembranes}, hep-th/9611203

[11] D. Garfinkel, G. T. Horowitz, A. Strominger, \textit{Charged black
holes in string theory},\ Phys. Rev. \textbf{D43 (}1991) 3140-3143

[12] J. Polchinski, Phys. Rev. Lett. \textbf{75 }(1995) 4724

[13] J. Polchinski, S. Chaudhuri, C. V. Johnson, \textit{Notes on D-Branes},
hep-th/9602052

[14] R. Jackiv, \textit{A particle field theorist's lectures on
supersymmetric, non-abelian fluid mechanics and d-branes}, physics/0010042

\end{document}